\documentclass{article}
\usepackage{arxiv}
\usepackage[utf8]{inputenc} 
\usepackage[T1]{fontenc}    
\usepackage{hyperref}       
\usepackage{url}            
\usepackage{booktabs}       
\usepackage{amsfonts}       
\usepackage{nicefrac}       
\usepackage{microtype}      
\usepackage{lipsum}
\usepackage{graphicx}
\usepackage{subfigure}
\usepackage{multirow}
\usepackage{caption}
\usepackage{cite}
\usepackage{amsmath,amssymb,amsfonts}
\usepackage{algorithmic}
\usepackage{graphicx}
\usepackage{textcomp}
\usepackage{xcolor}
\usepackage{subfigure}
\usepackage{pifont}
\newcommand{\cmark}{\ding{51}}%
\newcommand{\xmark}{\ding{55}}%
\def\BibTeX{{\rm B\kern-.05em{\sc i\kern-.025em b}\kern-.08em
    T\kern-.1667em\lower.7ex\hbox{E}\kern-.125emX}}
    
\title{MMCoVaR: Multimodal COVID-19 Vaccine Focused Data Repository for Fake News Detection and a Baseline Architecture for Classification\\
}

\author{
  Mingxuan Chen\thanks{This paper has been accepted for publication in ASONAM 2021.} \\
  Department of Electrical and Computer Engineering\\
  Stevens Institute of Technology\\
  Hoboken, NJ 07030 \\
  \texttt{mchen20@stevens.edu} \\
   \And
 Xinqiao Chu \\
  Department of Electrical and Computer Engineering\\
  Stevens Institute of Technology\\
  Hoboken, NJ 07030 \\
  \texttt{xchu2@stevens.edu}  \\
     \And
 K.P. Subbalakshmi \\
  Department of Electrical and Computer Engineering\\
  Stevens Institute of Technology\\
  Hoboken, NJ 07030 \\
  \texttt{ksubbala@stevens.edu}  \\
}

\begin{document}

\maketitle

\begin{abstract}
\label{sec:abs}

The outbreak of COVID-19 
has resulted in an ``infodemic" that has encouraged the propagation of misinformation about COVID-19 and cure methods which, in turn, could negatively affect the adoption of recommended public health measures in the larger population. 
In this paper, we provide a new multimodal (consisting of images, text and temporal information) labeled dataset containing news articles and tweets on the COVID-19 vaccine. We collected 2,593 news articles from 80 publishers for one year between Feb $16^{th}$ 2020 to May $8^{th}$ 2021 and 24184 Twitter posts (collected between April $17^{th}$ 2021 to May $8^{th}$ 2021).
We combine ratings from two news media ranking sites: Medias Bias Chart and Media Bias/Fact Check (MBFC) to
classify the news dataset into two levels of credibility: reliable and unreliable.  
The combination of two filters allows for higher precision of labeling. 
We also propose a stance detection mechanism to annotate tweets into three levels of credibility: reliable, unreliable and inconclusive.
We provide several statistics as well as other analytics like, publisher distribution, publication date distribution, topic analysis, etc.
We also provide a novel architecture that classifies the news data into misinformation or truth to provide a baseline performance for this dataset. We find that the proposed architecture has an F-Score of $0.919$ and accuracy of $0.882$ for fake news detection.
Furthermore, we provide benchmark performance for misinformation detection on tweet dataset.
This new multimodal dataset can be used in research on COVID-19 vaccine, including misinformation detection, influence of fake COVID-19 vaccine information, etc.
\end{abstract}

\section{Introduction}
The ongoing COVID-19 has been devasting in many ways. As of April $8^{th}$ 2021, the number of confirmed COVID-19 cases was $132,485,386$; including $2,875,672$ confirmed deaths \cite{WHO}.
This has caused panic about the possible economic fallout and recession. Along with the spread of the virus, misinformation about it has also spread virally. 
As the Director-General of the World Health Organization (WHO) said, fake news spreads faster and more easily than this virus.  
WHO defined the term, infodemics, as an excessive amount of information, including misinformation, disinformation and rumors that make it difficult to identify a solution, hampers effective public health response and creates confusion and distrust among people \cite{UN_misinfo}.
Therefore, it's necessary to track and analyze misinformation about this virus and the vaccines that have been developed for it. 

For e.g, a news item that claimed that the COVID-19 vaccine can alter DNA was circulated on the web \cite{hotEtal21}. 
Furthermore, the propagation of misinformation about vaccines affects the vaccination rates negatively, potentially leading to a worsening of the public health crisis. A study showed the number of Americans and Britons who would ``definitely'' get vaccinated dropped by 2.4 percent and 6.4 percent, respectively, after receiving misinformation about the COVID-19 vaccines \cite{looEtal20}.

\emph{This paper proposes a new Multimodal COVID-19 Vaccine Focused Data Repository (MMCoVaR) that includes 2,593 news articles and 24,184 related tweets with visual, textual and temporal information.}
Our work is inspired by \cite{ZhoEtal20} but with some differences which we will elaborate on later. 
The main contributions of our work are:
\begin{itemize} \itemsep 1pt
     \item An architecture for data collection from news sources and Twitter
     \item Annotation schemes for the 
     \begin{enumerate}
         \item news articles using two news media reliability ranking websites: (i) Medias Bias Chart \cite{Media_Bias_Chart} and (ii) Media Bias/Fact Check (MBFC) \cite{MBFC}, which allows for better accuracy of the annotation process;
     \item  Twitter data, using similarity based stance detection
     \end{enumerate} 
     
     
     \item Statistical and topic analysis for fake news articles and true news articles and a summary of the differences of topic distributions between them

    \item A novel attention based architecture for fake news classification which achieves an F-Score of 0.919 and accuracy of 0.882
    
    \item Several benchmark performances for misinformation detection for news and tweet data in this dataset
\end{itemize}

\section{Related Work}
\label{sec:related}
With the explosion of COVID-19, a lot of datasets have been released to support big data and deep learning based analyses of the disease and its effects.
One of the earliest COVID related datasets was the CORD-19 dataset \cite{WangEtal20} which is a resource of over 500,000 scholarly articles on COVID-19, SARS-CoV-2, and related coronaviruses. This dataset includes over 200,000 articles with full text.

Researchers have also mined social media to create COVID-19 related datasets.
Chakraborty et al \cite{ChaEtal20} released two types of Twitter-based datasets; one gathered from Dec 2019 to May 2020 and the other from Jan 2020 to Mar 2020. They provide the sentiment analysis for this tweet dataset.
The dataset in \cite{OrdEtal20} collects 23,830,322 tweets from Mar $24^{th}$ 2020 to Apr $9^{th}$ 2020 and utilizes Latent Dirichlet Allocation (LDA) and Uniform Manifold Approximation and Projection (UMAP) to analyze topics distribution. 
The COV19Tweets \cite{Lam20} is a Twitter-based 
dataset with more than 310 million COVID-19 specific English language tweets from Mar $20^{st}$ 2021 to Apr $17^{th}$ 2021. It analyzes the sentiment and geographical distribution of the tweets during the COVID-19 crisis and  proposes a sentiment scoring system. 
The CoronaVis \cite{KabEtal20} is also a Twitter dataset containing tweets gathered from US-based users from Mar $5^{th}$ 2020 to Jul $2^{nd}$ 2020. It focuses on sentiment and topic analysis. They also build a web application for topic trend analysis.
The GeoCoV19 \cite{QazEtal20} is a large-scale Twitter dataset containing more than 524 million multilingual tweets posted for 90 days from Feb $1^{st}$ 2020. GeoCoV19 focuses on geographical and topic distribution analysis and provides the distribution of topics by countries and cities.
The CoVaxxy \cite{DevEtal21} dataset is a Twitter dataset gathered between Jan $10^{st}$ 2021 to Feb $21^{th}$ 2021, focusing on geographical, topic and vaccine-related analysis. It provides a web information dashboard to track trending topics.
Among these datasets, only CoVaxxy provides COVID-19 vaccine-related analysis. 

While the above datasets focused on mainly topic, sentiment and other geographic properties of the tweets; several researchers have also started to work on COVID-19 misinformation related datasets. We discuss these in the following subsection.

\subsection{Misinformation Focused Datasets}
\label{subsec:misinfo_data}

The CMU-MisCov19 dataset \cite{MemEtal20}  contains 4,573 manually labeled tweets with 17 categories: ``Irrelevant", ``Conspiracy", ``True Treatment", ``True Prevention", ``Fake Cure", ``Fake Treatment", ``False Fact or Prevention", ``Correction/Calling out", ``Sarcasm/Satire", ``True Public Health Response", ``False Public Health Response", ``Politics", ``Ambiguous/Difficult to Classify", ``Commercial Activity or Promotion", ``Emergency Response", ``News" and ``Panic Buying".
The analysis in this work draws the conclusion that misinformed communities are denser and more organized than informed communities, a large majority of misinformed users may be anti-vaxxers and that informed users tend to use more narratives than misinformed users when making up their minds.

The Constraint@AAAI2021 dataset \cite{PatEtal20} extracts 10,700 social media posts from Facebook, Twitter, Fact Checking, etc. It is also a manually labeled dataset with two categories: real and fake. It provides benchmark performances using four baseline models: (1) Decision Tree, (2) Logistic Regression, (3) Gradient Boost and (4) Support Vector Machine (SVM). The best performance is 93.46\% on the F1-score using SVM.

The COVIDLIES dataset \cite{HosEtal20} identifies
86 common misconceptions using
Wikipedia articles on COVID-19 related misconceptions.
The dataset contains 6,761 tweets, identified and human annotated by researchers. 
The dataset creates misconception-tweet pairs with three labels: ``Agree", ``Disagree" and ``No Stance" which describes whether the tweet agrees or disagrees with the misconception or has no stance with respect to the misconception. The authors use BERTScore \cite{ZhaEtal19} to compute a similarity metric on tweet-misconception pairs and use that for stance detection.

ReCoVery \cite{ZhoEtal20} collects 2,029 COVID-19 related news articles and 140,820 tweets containing
textual, visual, temporal, and network information from Jan $21^{th}$ 2020 to May $26^{th}$ 2020. The authors then use two news quality rating websites to annotate the news articles into two categories: reliable news and unreliable news. 
They also include the performance of four baseline 
models for fake reliable/unreliable news detection as benchmarks: (1) LIWC+DT, (2) RST+DT, (3) Text-CNN and (4) SAFE a neural network based method. 
%
%

Although CMU-MisCov19~\cite{MemEtal20}, Constraint@AAAI2021~\cite{PatEtal20}, COVIDLIES~\cite{HosEtal20} and ReCoVery~\cite{ZhoEtal20} 
provide labeled datasets, 
our dataset adds some additional value. We list the key differences below:
\begin{itemize}
\item The CMU-MisCov19 with 17 categories doesn't provide any benchmark performance for misinformation detection, is a manually labeled dataset and contains fewer tweets than our dataset. 
\item The CMU-MisCov19, Constraint@AAAI2021 and the COVIDLIES datasets only provide labeled tweets dataset and not the relevant news articles. These datasets essentially do not concern themselves with the news article-tweet pairing.
\item The ReCoVery dataset is comprehensive, consisting of news articles and relevant tweets. However, the tweets are annotated as misinformation or true by simply inheriting the label of the relevant news articles. That is, there is no stance detection involved, which might mean that a tweet that refers to a fake news article but calls it fake can be tagged as a fake tweet and vice versa. This paper also does not provide fake tweet detection benchmarks.
\item The COVIDLIES dataset uses BERTScore to compute a similarity metric on tweet-misconception pairs and uses this as a stance detection mechanism. They do not use news articles. We propose a method to detect the stance of a tweet with respect to the news article and then use that to label the tweet. More details on our stance detection approach can be found in Sec.~\ref{SubSec:social_collect}.
\item Finally, our dataset is focused specifically on misinformation in the realm of COVID-19 vaccination.
\end{itemize}
%
We will elaborate more on the differences between these datasets and the proposed dataset in Section~\ref{SubSec:comp}. In short, we provide a Multimodal COVID-19 Vaccine Focused Data Repository (MMCoVaR). We annotate the news dataset using two news website source checking filters and the tweets dataset based on stance detection. We also provide benchmark performances on news and tweets datasets for further comparison.

\section{Data}
\begin{figure*}[htbp]
\includegraphics[width=1\textwidth]{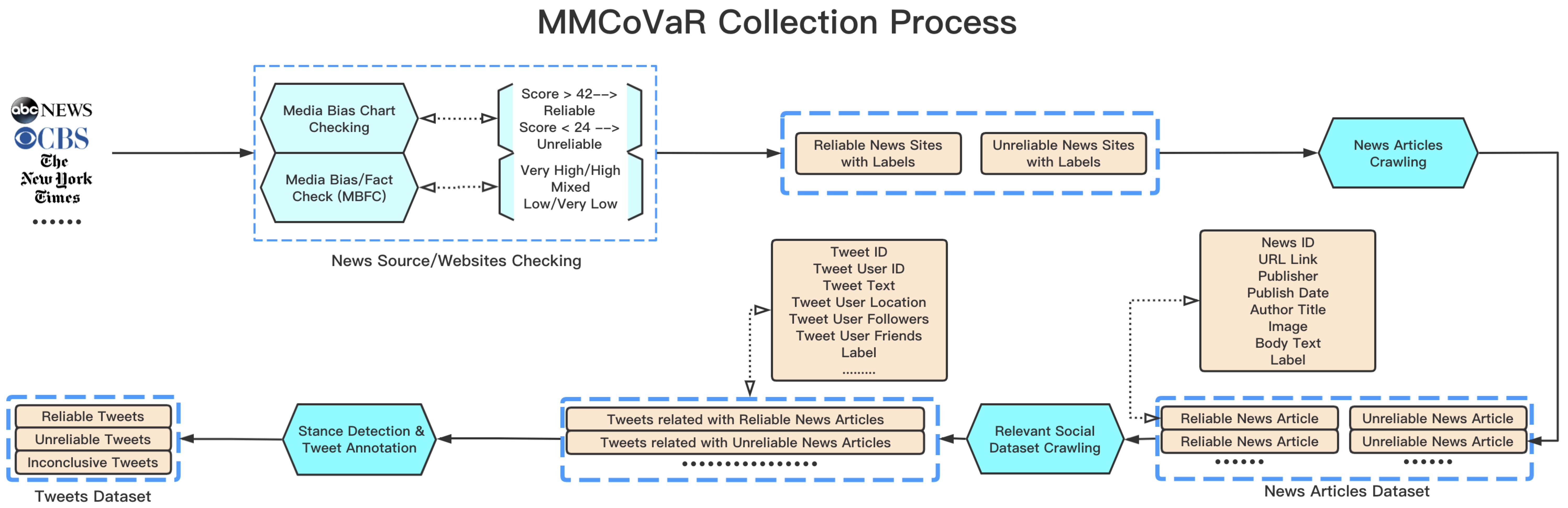}
\caption{Data collection process including news dataset and tweet dataset. The news dataset is validated using two sources of credibility. 
The tweets mentioning URLs of news articles are checked for stance and then labeled as reliable, unreliable or inconclusive.} \label{fig_collect}
\end{figure*}

As mentioned earlier, our goal is to provide a labeled dataset that comprises of news articles as well as social media posts (tweets) to 
support research in COVID-19 vaccine related misinformation. 
We detail how we collect and annotate the news data in Section~\ref{SubSec:News Data Collection} and social media data in Section~\ref{SubSec:social_collect}. Our dataset is compared with other datasets in Section~\ref{SubSec:comp}. We also provide some statistics and topic analysis in Section~\ref{SubSec:Data Sta Analysis} and Section~\ref{SubSec:Data Top Analysis}.
Figure~\ref{fig_collect} shows a schematic overview of the proposed data collection and annotation process, for both news articles and social media.

\subsection{News Data Collection and Annotation}\label{SubSec:News Data Collection}
%

Media Bias Chart Checking \cite{Media_Bias_Chart} provides 120 websites and Media Bias/Fact Check (MBFC) \cite{MBFC} provides around 3700 websites with their ratings. Of these, 80 websites are common between Media Bias Chart Checking and Media Bias/Fact Check (MBFC). 
As shown in Fig~\ref{fig_collect}, we classify these 80 websites into two groups according to our ranking system described below.


Media Bias Chart is a website reliability visualization tool to display fact checking information and political bias information of news sites, which are annotated by their analyzers. 
The reliability score from Media Bias Chart is between $0$ to $64$; $0$ indicates
lowest credibility, and $64$ indicates highest credibility.
Media Bias/Fact Check (MBFC) categorizes data
into six levels of reliability using manual fact checking 
by their analyzers. These levels are: 
``Very High", ``High", ``Most Factual", ``Mixed", ``Low" and ``Very Low".
%
For our purposes, to improve the precision, we discarded the websites whose reliability levels from Media Bias/Fact Check (MBFC) are labeled ``Most Factual" or ``Mixed". 
We use the following criteria to label the remaining news articles: 

\begin{itemize}
    \item Label $\leftarrow$ \emph{Reliable News}\\ If  MBFC ranking is `` High" or ``Very High"\\ \textbf{AND} Media bias chart score is greater than $42$
    
    \item Label $\leftarrow$ \emph{Unreliable News}\\ If MBFC factual ranking is `` Low" or ``Very Low"\\\textbf{AND} Media bias chart score is below $24$.
\end{itemize}

Then we use newspaper library \cite{newspaper}, a news articles extraction tool, to crawl COVID-19 vaccine related news articles using the following nine keywords:
\textit{``vaccine", ``vaccinated", ``COVID-19 Pfizer", ``COVID-19 Moderna", ``COVID-19 Janssen", ``Moderna vaccines", ``Janssen vaccines", ``Johnson \& Johnson’s vaccines", ``Biontech vaccine"}.

For the news dataset, we provide the following information:
\textit{``News ID", ``URL Link", ``Publisher", ``Publish Date", ``Author Title", ``Image", ``Body Text", ``Label"}.
In total, the dataset contains 2,593 annotated news articles consisting of 958 unreliable news articles and 1,635 reliable news articles as shown in Table~\ref{tab_data}.
We also provide the following statistics for the news articles: (i) news publisher distribution, (ii) authors distribution (number of articles  written by a single author and multiple authors), 
(iii) news words distribution, etc. in the Section~\ref{SubSec:Data Sta Analysis} and topic analysis in the Section~\ref{SubSec:Data Top Analysis}.

\subsection{Social Media Data Collection and Annotation}
\label{SubSec:social_collect}

\begin{figure}[htbp]
    \centering{
    \includegraphics[width=0.36\textwidth]{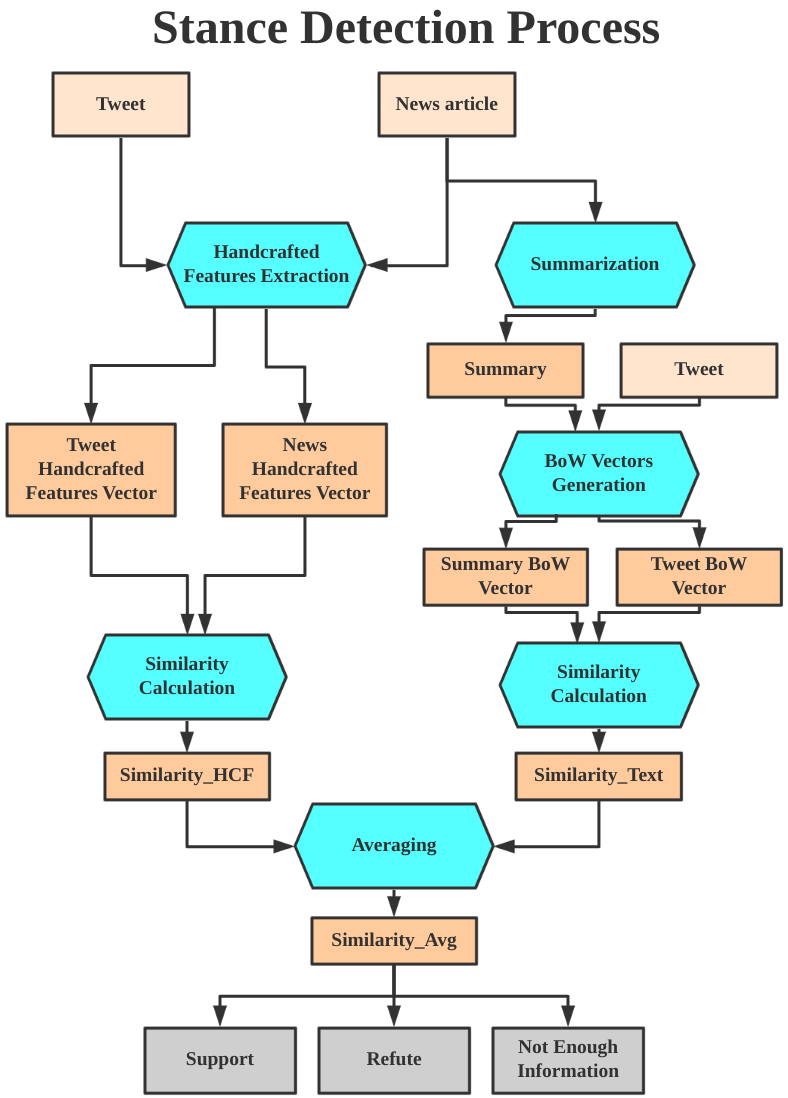}
    }
    \caption{Stance Detection Process}
    \label{fig:stance}
\end{figure}

Using the Twitter developer API \cite{Twitter_deve}, we 
crawl all tweets that mention the URL
of the news articles in the news dataset that we just
created. Tweets were gathered from Apr $16^{th}$ 2021 to May $7^{th}$ 2021.
%
As mentioned earlier a mere mention of the URL of an unreliable news article does not automatically 
mean that the tweet is also fake. A user could be agreeing or disagreeing with the content of the news in the URL. Hence, we need to determine the stance that the tweet takes with respect to the news article mentioned in the URL.

\subsubsection{Stance Detection}
The proposed stance detection process is illustrated in the Fig~\ref{fig:stance}.
Recently it was shown that lexical features and readability features can be highly indicative of the stance of a written text \cite{HanEtal18}. This work used type-token-ratio (TTR), the measure of textual lexical diversity (MTLD), automated readability index (ARI), Flesch-Kincaid grade (FKG) and Flesch reading ease (FRE) and some LSTM algorithms to detect stance.

In our approach, we use all of the above features, and add root type-token ratio (RTTR), corrected type-token ratio (CTTR) and Dale Chall readability score (DCR)~\cite{Mcc87} to create a handcrafted feature vector: ${\rm HCF}_{\rm news}$ and ${\rm HCF}_{\rm tweet}$, where the ${\rm HCF}_{\star} = [{\rm TTR}_{\star},{\rm RTTR}_{\star},{\rm CTTR}_{\star},{\rm MTLD}_{\star},{\rm ARI}_{\star},{\rm FKG}_{\star},{\rm FRE}_{\star},{\rm DCR}_{\star}]$.
One handcrafted feature vector is created for each news article and the tweet that mentions it.
Then, we compute the similarity, ${\rm Similarity}_{\rm HCF}$,  between the two handcrafted feature vectors using cosine similarity metric~\cite{GunEtal18}.
We then compute the summary of the article. 
We use the summary of the article instead of the entire article in order to be closer to the tweet length which is only $280$ characters long.
%
We create a BoW vector for the summary of each news article and every tweet. 
Then, we estimate the similarity, ${\rm Similarity}_{\rm text}$, between every tweet and news article pair by computing the cosine similarity between the BoW vectors for each of these.
The average of ${\rm Similarity}_{\rm text}$ and ${\rm Similarity}_{\rm HCF}$, $\rm {Similarity}_{\rm avg}$, is used to detect the stance of the tweet with respect to the news article.
$\rm {Similarity}_{\rm avg}$ is a real number in the range $[0, 1.0]$, where $0$ means least similarity and $1$ means highest similarity.
%
In our case, we set the stance as ``support" when the score is in range $[0.6,1]$, ``refute" when the score is in range $[0,0.4]$ and ``Not Enough Information" when the score is in range $[0.4,0.6]$. 
Then we annotate each tweet based on the credibility of the news articles and the stance of social media as follows:
\begin{itemize} \itemsep -2pt
    \item Label $\leftarrow$ \emph{Reliable}\\ If the relevant news is reliable and stance is ``Support" \textbf{OR} the relevant news is unreliable and stance is ``Refute"
    \item Label $\leftarrow$ \emph{Inconclusive}\\ If the relevant news is reliable and stance is ``Not Enough Information" \textbf{OR} the relevant news is unreliable and stance is ``Not Enough Information"
    \item Label $\leftarrow$ \emph{Unreliable}\\ If the relevant news is reliable and stance is ``Refute" \textbf{OR} the relevant news is unreliable and stance is ``Support"    
\end{itemize}

Based on the Twitter development API agreement and policy \cite{Twitter_policy}, we can only provide the tweet ID of the Twitter data. Additional information about the tweet can be obtained by twarc \cite{twarc} or hydrator \cite{hydrator}. 
%
The dataset can be downloaded at \cite{our-misinfo-data} \footnote{Our collection relies upon publicly available data and is registered as IRB exempt by Stevens Institute of Technology IRB (approved protocol 2021-035(N)). We release the data set with the stipulation that those who use it must comply with Twitter’s Terms and Conditions.}. 
In total, we extracted 24,184 tweets, including 3,092 unreliable tweets, 17,234 inconclusive tweets and 3,858 reliable tweets. The detail information of annotated news articles and related tweets is shown in Table~\ref{tab_data}.

\begin{table}[htbp]
\begin{center}
\begin{tabular}{llll}
\hline
\hline
\multicolumn{4}{l}{{\bfseries News Articles Data}}\\
\hline
Credibility &Label\qquad\qquad &Amount of News & Total Amount\\
\hline
Unreliable & 0 & 958 & {2593} \\
\cline{1-3}
Reliable&1 & 1635 & ~ \\
\hline
\hline
\multicolumn{4}{l}{{\bfseries Tweets Data}}\\
\hline
Credibility &Label&Amount of Tweet& Total Amount\\
\hline
Unreliable & 0 & 3092 &  ~\\
\cline{1-3}
Inconclusive &1 & 17234 & {24184}\\
\cline{1-3}
Reliable&2 & 3858 &~ \\
\hline
\hline
\end{tabular}
\end{center}
\caption{Data Statistics}
\label{tab_data}
\end{table}


\begin{table*}[htbp]
\begin{center}
\begin{tabular}{llllll}
\hline
\hline
\textbf{Dataset Name} & \textbf{MB Chart}  & \textbf{MBFC}& \textbf{Manually Check}&\textbf{Stance Detection}\\
\hline
CMU-MisCov19\cite{MemEtal20} & \xmark & \xmark& \cmark&\xmark\\
\hline
Constraint@AAAI2021\cite{PatEtal20} & \xmark  & \xmark& \cmark&\xmark\\
\hline
COVIDLIES\cite{HosEtal20} & \xmark & \xmark& \cmark&BERTScore\\
\hline
ReCOVery\cite{ZhoEtal20} & \xmark  & \cmark& \xmark&\xmark\\
\hline
{\bfseries MMCoVaR} & \cmark  & \cmark& \xmark&Lexical,Readability,Language Embedding \\
\hline
\hline
\end{tabular}
\end{center}
\caption{Comparison of collection process among different misinformation focused datasets. The \cmark and \xmark means the dataset used the corresponding approach and not, respectively. In this table, MB Chart represents Media Bias Chart, MBFC represents Media Bias/Fact Check and Manually represents manual annotation. Stance Detection column captures whether the Twitter dataset uses stance detection to label tweets that mention news article URLs: only COVIDLIES and MMCoVaR use stance detection for tweet annotation. 
}
\label{tab:data_collect}
\end{table*}
\begin{table*}[htbp]
\begin{center}
\begin{tabular}{llllll}
\hline
\hline
\textbf{Dataset Name} & \textbf{News Data} & \textbf{News Image}& \textbf{Social Data} & \textbf{Vaccine} & \textbf{Time Span}  \\
\hline
CMU-MisCov19\cite{MemEtal20}&\xmark& \xmark  & \cmark & \xmark& Not available \\
\hline
Constraint@AAAI2021\cite{PatEtal20} &\xmark& \xmark  & \cmark & \xmark& Not available \\
\hline
COVIDLIES\cite{HosEtal20} &\xmark& \xmark & \cmark& \xmark & Not available \\
\hline
ReCOVery\cite{ZhoEtal20} &\cmark& \cmark & \cmark& \xmark & 1/21/2020 to 5/26/2020 \\
\hline
{\bfseries MMCoVaR} &\cmark& \cmark  & \cmark & \cmark & 2/16/2020 to 3/17/2021  \\
\hline
\hline
\end{tabular}
\end{center}
\caption{Comparison of the properties of different misinformation focused datasets. The \cmark and \xmark means the dataset contains the specific information or not.
}
\label{tab:data_property}
\end{table*}
\subsection{Comparison with other Datasets}
\label{SubSec:comp}
We compare the collection process and properties of our dataset with other recent COVID-19 related misinformation datasets in Table~\ref{tab:data_collect} and Table~\ref{tab:data_property}.

The comparison of data collection process is shown in Table~\ref{tab:data_collect}. As mentioned in  Section~\ref{sec:related}, only CMU-MisCov19, Constraint@AAAI2021, COVIDLIES and ReCoVery 
datasets provide misinformation-annotated data, the rest of the datasets are not focused on misinformation and hence do not label for misinformation or truth.
%
CMU-MisCov19\cite{MemEtal20} and Constraint@AAAI2021\cite{PatEtal20} use manual annotation without any web site reliability checking. ReCoVery\cite{ZhoEtal20} utilizes two news quality rating websites to label for misinformation after the news data and tweets have been gathered. 
Our proposed dataset, MMCoVaR, merges the MB Chart and MBFC to check for news site reliability. Among all the datasets, only COVIDLIES and our proposed MMCoVaR use stance detection for annotating tweets. While COVIDLIES uses BERTScore to compute similarity between a tweet and misconception pair; we use stance detection to check whether the tweet agrees with the news article or not.

The properties of the different datasets is shown in Table~\ref{tab:data_property}. All datasets provide social media information; however, only the ReCoVery \cite{ZhoEtal20} and the proposed MMCoVaR dataset provide the relevant news articles information. Also, only CoVaxxy \cite{DevEtal21} and MMCoVaR provide COVID-19 vaccine related data. However, the CoVaxxy focuses on the topic analysis, geographical distribution, etc. Therefore, they do not label this dataset for misinformation.
Furthermore, the time span of the dataset presented in this paper MMCoVaR is the longest one among all datasets.

\subsection{Data Statistics}\label{SubSec:Data Sta Analysis}
In this subsection, we provide some additional statistics about our dataset, which may add some insight into misinformation surrounding COVID-19 vaccinations.
\begin{figure}[htbp]
    \centering{
    \includegraphics[width=0.5\textwidth]{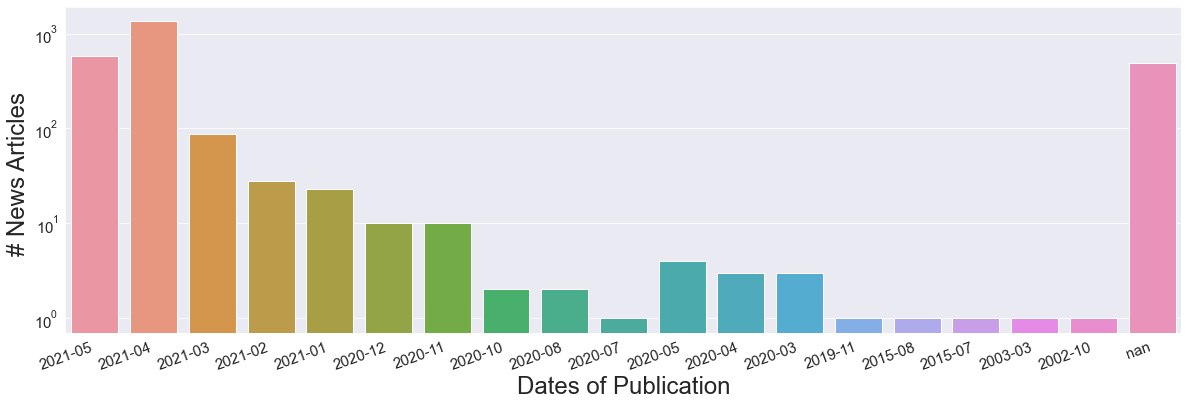}
    }
    \caption{Publication Date Distribution}
    \label{fig:publish_Date}
\end{figure}
%
\begin{figure}[htbp]
    \subfigure[Publisher Distribution on Reliable News]{
        \includegraphics[width=0.5\textwidth]{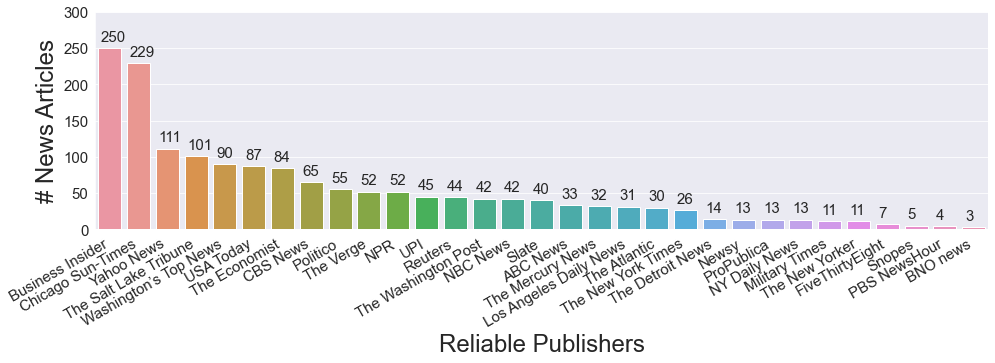}
    }
    \subfigure[Publisher Distribution on Unreliable News]{
	\includegraphics[width=0.5\textwidth]{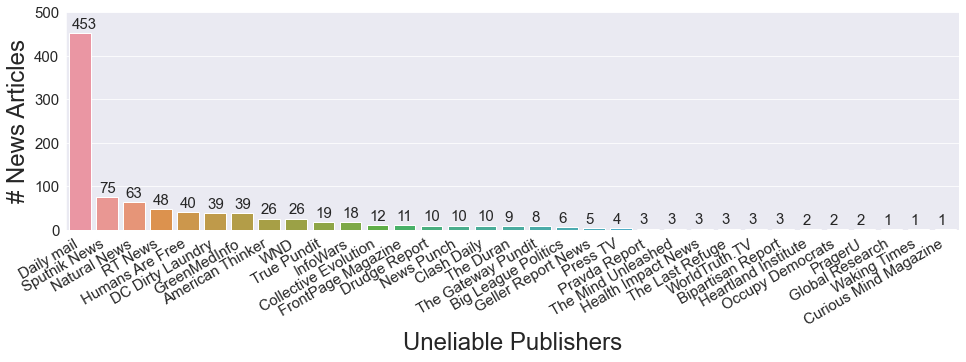}
    }
    \caption{News Publisher Distribution}
    \label{fig:News_publiser}
\end{figure}
%
\begin{figure}[htbp]
    \centering
    \subfigure[Distribution on Reliable News]{
        \includegraphics[width=0.2\textwidth]{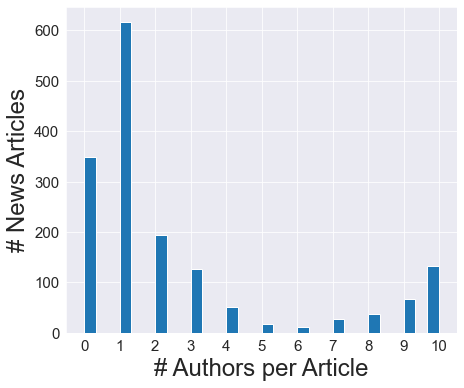}
    }
    \subfigure[Distribution on Unreliable News]{
	\includegraphics[width=0.2\textwidth]{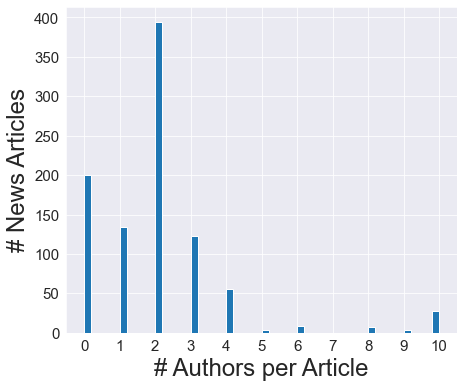}
    }
    \caption{News Author Distribution}
    \label{fig:News_Author}
\end{figure}
%
\begin{figure}[htbp]
    \centering
    \subfigure[Distribution on Reliable News]{
        \includegraphics[width=0.2\textwidth]{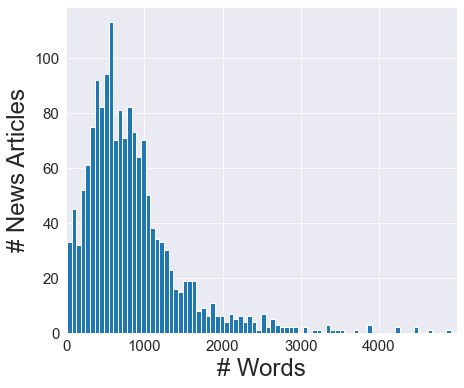}
    }
    \subfigure[Distribution on Unreliable News]{
	\includegraphics[width=0.2\textwidth]{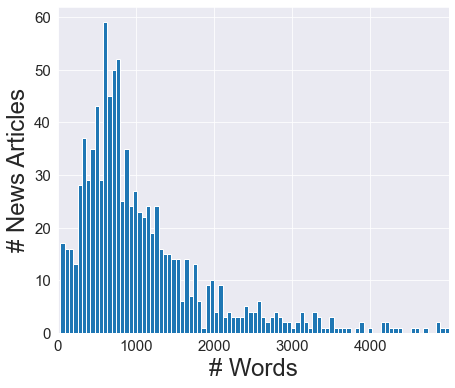}
    }
    \caption{Word Distribution for News Articles}
    \label{fig:News_Words}
\end{figure}

{\bfseries Publication Date Distribution:} The distribution of the publication date is shown in Fig~\ref{fig:publish_Date}. From this figure we see
that the there is a steep uptick in the news articles about COVID-19 vaccines since Nov 2020.
This timeline correlates with the fact that the COVID-19 vaccine became available in November and Pfizer COVID-19 vaccine was approved in the UK in Dec $2^{\rm nd}$ \cite{Pfizer_authorisation}.

{\bfseries News Publisher Distribution:}
Fig~\ref{fig:News_publiser} shows the number of articles published
by different news media outlets in our dataset.
Fig~\ref{fig:News_publiser} (a) shows the distribution of reliable publishers which is dominated by ``Business Insider" and ``Chicago Sun-Times". 
Fig~\ref{fig:News_publiser} (b) shows the distribution of articles by unreliable publishers and is dominated by the``Daily Mail" which contributed 463 articles, accounting for $48.3\%$ of the  unreliable news. 

{\bfseries News Author Distribution:} We provide the distribution of the number of authors for each news article in Fig~\ref{fig:News_Author}. 
We note that the number of authors for most news articles is fewer than 5; however, reliable news articles are dominated by single author articles, whereas unreliable news articles are dominated by two author articles.

{\bfseries Word Distribution for News Articles:}
We plot the word count distribution for news articles for both reliable and unreliable news in Fig~\ref{fig:News_Words}. The mean, median and mode of word number for both cases are: 
911, 708 and 645, respectively for reliable news; 1169, 812 and 744, respectively for unreliable news.
The similarity of the distribution of word count for reliable and unreliable news may indicate that the length of these articles may not be a useful 
feature to distinguish between real and fake news.

\subsection{Data Topic Analysis}\label{SubSec:Data Top Analysis}
\begin{figure}[htbp]
    \centering
    \subfigure[Reliable News Distribution]{
        \includegraphics[width=0.22\textwidth]{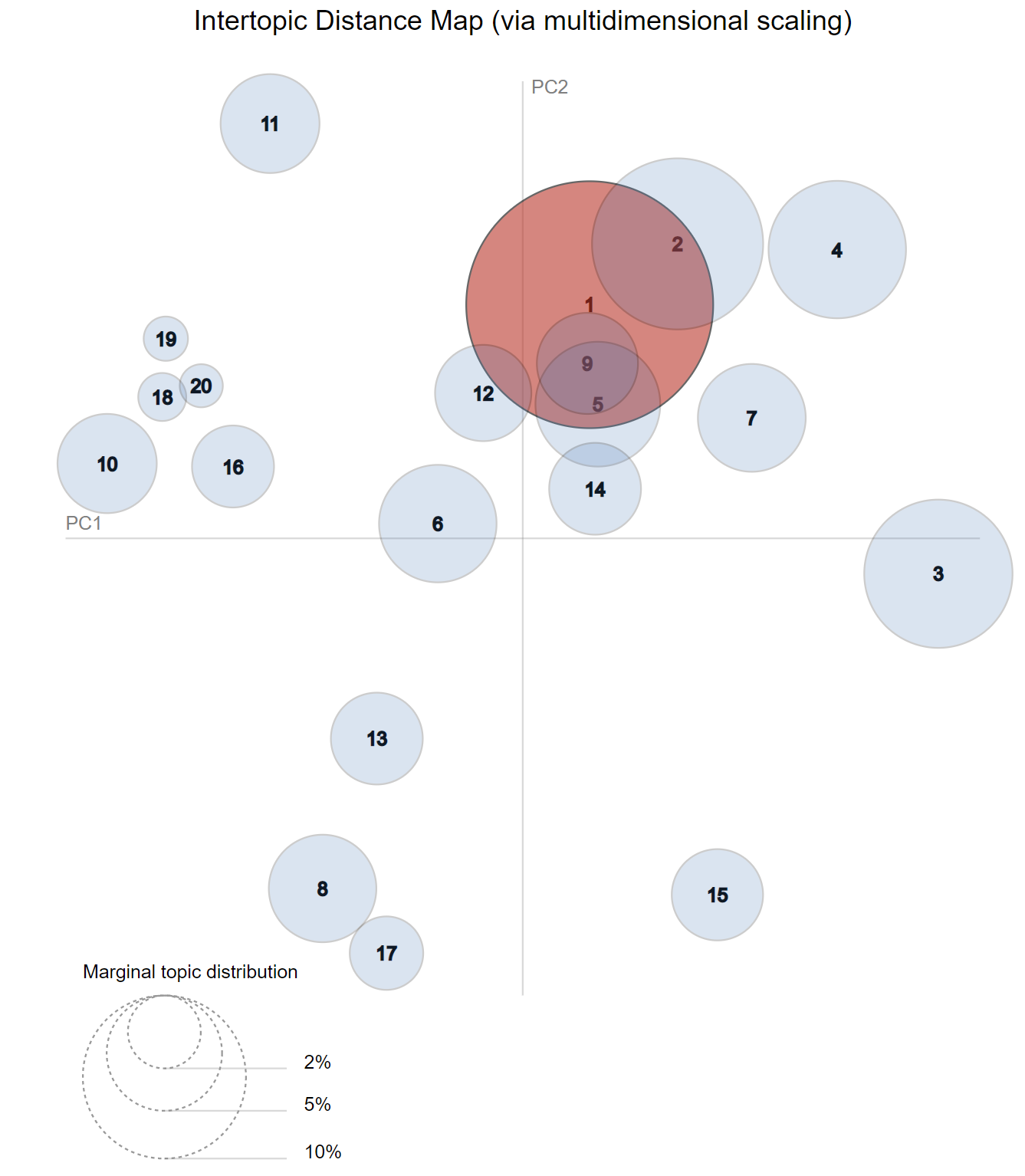}
    }
    \subfigure[Unreliable News Distribution]{
	\includegraphics[width=0.22\textwidth]{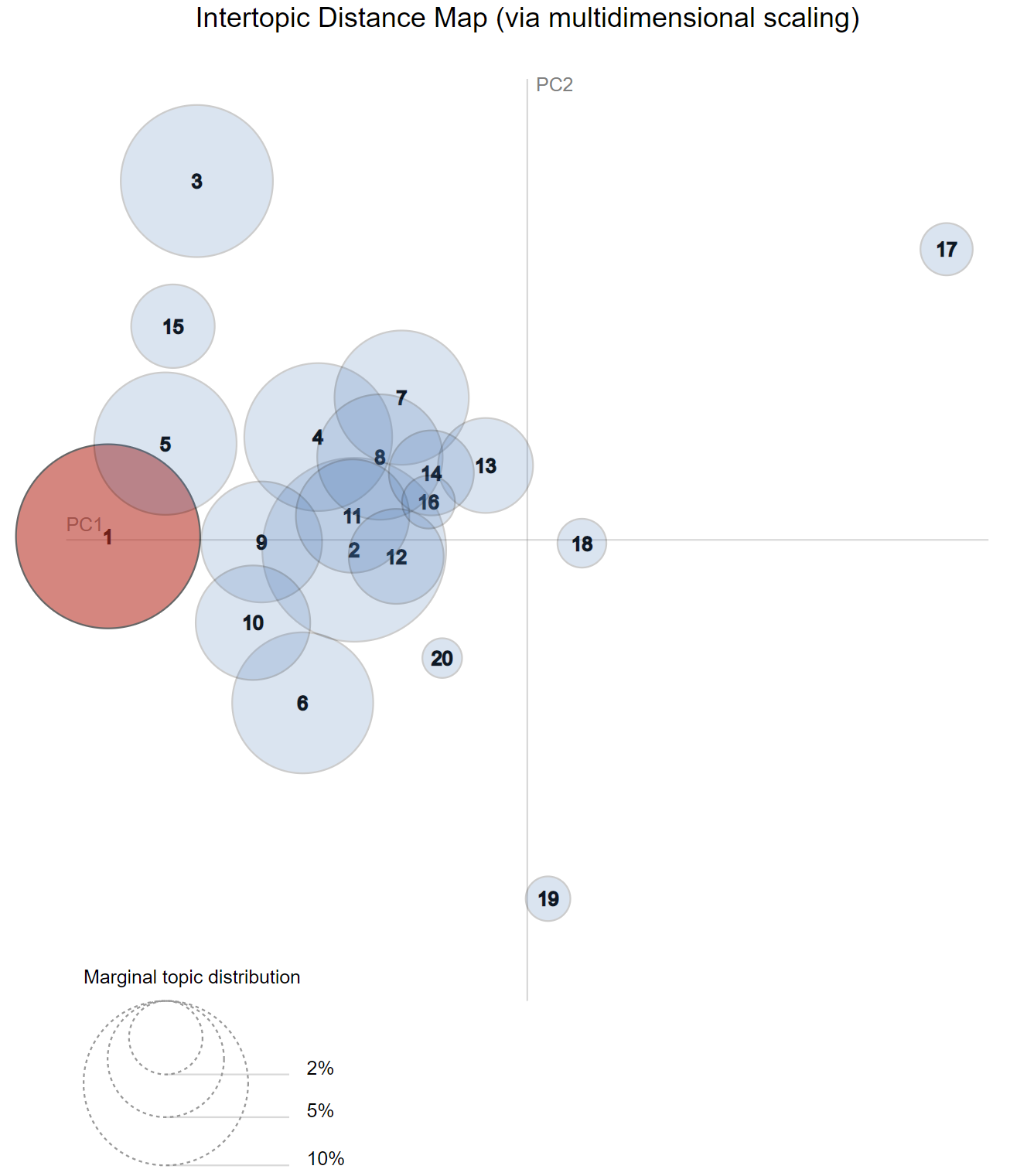}
    }
    \caption{LDA Cluster of Reliable and Unreliable News}
    \label{fig:LDA}
\end{figure}

Topic analysis is a useful tool to gather insights
into the kinds of topics that are discussed in social media 
in all three categories of tweets (unreliable, reliable and inconclusive).
In this section, we utilize Latent Dirichlet Allocation (LDA)\cite{bleEtal03} 
to analyze the distribution of topics in our data repository, since it is an unsupervised method and more suitable for our dataset. 

We extract 20 different topics using LDA analysis. Each topic was described using the 30 most relevant words for that topic. A visual representation of these 20 topics is shown in Fig~\ref{fig:LDA}. 
Each topic is represented by a circle and the size of the circle is proportionate to the number of news articles in the dataset that covers that topic.
We find that distribution of topics under the reliable news category is more dispersed than that under unreliable news category.

\section{Proposed Architecture for Misinformation Detection}
\label{sec:propose}

\begin{figure}[htbp]
\centering\includegraphics[width=0.4\textwidth]{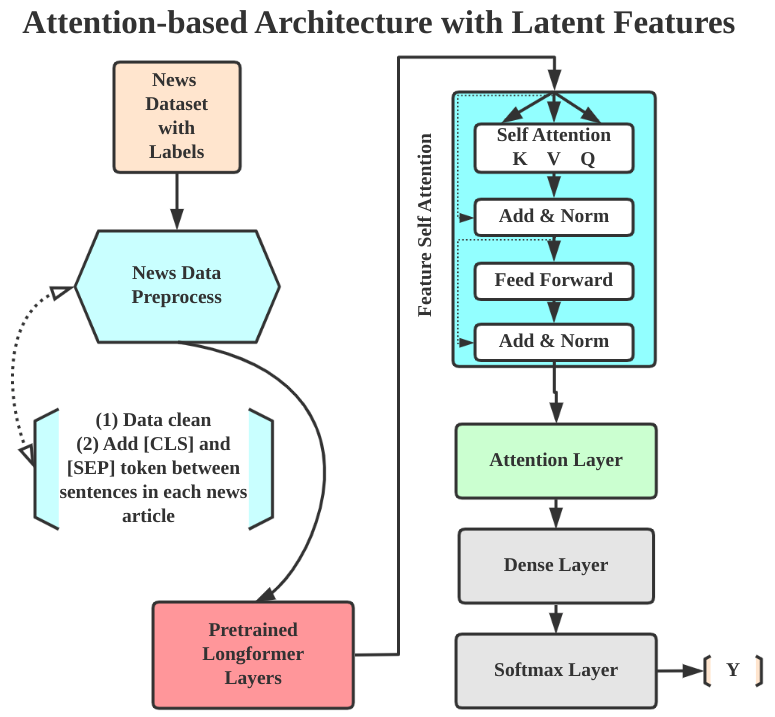}
\caption{The architecture with latent features using Longformer based sentence embedding.} \label{fig_model}
\end{figure}
We propose an attention based architecture that uses Longformer \cite{BelEtal20}
to embed the news articles. The Longformer can capture the long-range dependencies in text to create a fixed dimension embedding of the sentence/article. 

We preprocess the data by removing the website links, 
stop words and non-alphanumeric characters from the text. Then we add special tokens like [CLS] and [SEP] between sentences. 
Let ${N}_{i}$ be the ${i}^{\rm {th}}$ preprocessed news article in the news dataset.
%
Let $n_i$ be the language embedding
of the ${i}^{\rm {th}}$ preprocessed news article, $N_i$, obtained using 
Longformer. $n_i$ is a $768$-dimension vector.
The resulting vectors, ${n}_{i}$, are used to construct the matrix ${M}=\{{n}_{1},{n}_{2}, \ldots,{n}_{m}\}$. Subsequently, we feed this matrix ${M}$ into an ${h}$-layer Multi Head Attention (MHA) module~\cite{vasEtal17}. The MHA is followed by an attention layer, a dense layer and a softmax layer to get the final classification.
The architecture is depicted in Fig~\ref{fig_model} and is inspired by our earlier work on interpretable fake tweet detection~\cite{CheEtal20}.

\section{Experiment}

\subsection{Experiment Setup}
\label{sec:setup}
We implement all the models in Pytorch and 
train them to minimize the cross-entropy loss function of predicting the class label of tweets in the training set. 
For all models with attention architecture in our 
experiments, we use a $6$-layer multi-head attention (MHA) module and the stochastic 
gradient descent (SGD) \cite{Rud16} as the optimizer for training.
Moreover, we split the dataset in the $4:1$ ratio for
training and testing and use $5$-fold cross validation to prevent over-fitting the model.

Before we discuss the models for misinformation detection, we first describe the handcrafted features that we use in some of these models.

\subsection{Handcrafted Features}
Handcrafted features can play an important role in misinformation detection as demonstrated in \cite{CheEtal20}. 
We extract the following handcrafted features: $37$ PoS tags and one sentiment feature (average value of words' sentiment polarity) using NLTK \cite{LopEtal02}, four psycho-linguistic features (`FamiliarityScore', `ConcretenessScore', `ImagabilityScore' and `AgeofAcquisitionScore'), 
four vocabulary richness features (Honore’s Statistic (HS), Sichel Measure (SICH), Brunet’s  Measure (BM) and Text-Type Ratio (TTR)) and two readability features (Automated ReadabilityIndex (ARI) and Flesch-Kincaid readability (FKR) scores) \cite{FraEtal16}.
Examples of some of these PoS tags are shown in Table~\ref{tab:postags}.

\subsection{Models}
\label{sec:results}
We provide the performance (in terms of accuracy, F1 score, precision and recall) of several benchmark architectures on our dataset. The comparisons are shown in Table~\ref{tab_news_res} and Table~\ref{tab_tweets_res}.
The benchmark architectures are described below.
\begin{itemize}
\itemsep -2pt
\item \textbf{Logistic Regression \cite{HosmerEtal13}:} Logistic regression models the probabilities for classification problems with two possible outcomes.
\item \textbf{Gradient Boost \cite{scikit-learn}:} Gradient Boost is an additive model in a forward stage-wise fashion for classification.
\item \textbf{Decision Tree\cite{SafEtal91}:} Decision tree is a non-parametric supervised learning method used for classification.
\item \textbf{Handcrafted Features+Attention (HCF+Att) \cite{CheEtal20}:} An attention based architecture which uses handcrafted features (HCF) for fake news detection. It can also provide explanations in a classification task.
\item \textbf{Handcrafted Features+LSTM (HCF+LSTM):} A basic 1-layer LSTM architecture which uses handcrafted features (HCF) for the fake news detection task. 
\item \textbf{Proposed Architecture: Latent Features using \\Longformer based Sentence Embedding+Attention\\(LFLSE+Att):} 
LFLSE+Att is the attention based architecture we propose in this work and described in Section~\ref{sec:propose}. 
\end{itemize}

\subsection{Performance Analysis}
\label{sec:perf-anal}
\begin{table}
\begin{center}
\begin{tabular}{lllll}
\hline
\hline
Method & Accuracy \qquad& F Score \qquad & Precision\qquad & Recall\qquad \\
\hline
Logistic Regression & 0.861 & 0.869 &\textbf{0.906}  &0.861 \\
\hline
Gradient Boost & 0.880 & 0.883  &0.897  &0.880\\
\hline
Decision Tree& 0.815 & 0.815  &0.816  &0.815\\
\hline
HCF+Att&0.679&0.808&0.688&\textbf{0.979}\\
\hline 
HCF+LSTM&0.685&0.799&0.714&0.909\\
\hline
\textbf{LFLSE+Att}&\textbf{0.882}&\textbf{0.919}&0.874&0.970\\
\hline
\hline
\end{tabular}
\end{center}
\caption{Misinformation Detection on News Dataset}\label{tab_news_res}
\end{table}

\begin{table}
\begin{center}
\begin{tabular}{lllll}
\hline
\hline
Method & Accuracy \qquad& F Score \qquad & Precision\qquad & Recall\qquad \\
\hline
Logistic Regression & 0.839 & \textbf{0.858} & 0.908 &\textbf{0.839}\\
\hline
Gradient Boost & 0.748 & 0.820  &\textbf{0.950}  &0.748\\
\hline
Decision Tree & \textbf{0.857} & 0.857  & 0.860 & 0.857\\
\hline
\hline
\end{tabular}
\end{center}
\caption{Misinformation Detection on Tweet Dataset}\label{tab_tweets_res}
\end{table}
The classification results on news dataset are shown in Table~\ref{tab_news_res}. 
We can see that 
the latent features extracted using Longformer based Sentence Embedding-Attention (LFLSE-Att) performs best compared with the other architectures on the MMCoVaR news dataset using the same experiment setup 
with an accuracy and F-Score of 0.882 and 0.919, respectively. The performance of LFLSE-Att in terms of recall is 0.970, which is close to the best score (0.979).
This points to the fact that features extracted from the Longformer are potentially more useful compared to handcrafted features for classifying news into fake or real.
In addition, the performance of Handcrafted Features-Attention (HCF-Att) in terms of recall is better than the other models, while Logistic Regression performs best in terms of precision.
The baseline performances of the classification we provided here can be used for comparison with other models in the future.
Furthermore, more classes of features including visual information and social media information can be added into our architecture, if desired, to construct a composite classifier engine. 

The classification results on tweets are shown in Table~\ref{tab_tweets_res}. Here, logistic regression outperforms other methods in terms of F-Score and Recall. 

\begin{table}[ht]
\begin{center}
\begin{tabular}{p{1.8cm}c} 
\hline
\hline
\textbf{Tag} & \textbf{Description}  \\ 
\hline
SYM&Symbol\\
RB&Adverb\\
CD&Cardinal number\\
JJ&Adjective\\
VBZ&Verb,3rd person singular present\\
MD&Modal\\
PRP&Personal pronoun\\
NNP&Proper noun, singular\\
NNS&Noun, plural\\
WRB&Wh-adverb\\
CC&Coordinating conjunction\\
VBG&Verb, gerund or present participle\\
VB&Verb, base form\\
...& ...\\
\hline
\hline
\end{tabular}
\caption{Examples of PoS tags feature}
\label{tab:postags}
\end{center}
\end{table}

\label{sec:experiment}
\section{Conclusion and Future}
\label{sec:conclusion}
The ``infodemic'' caused by the COVID-19 has sparked the spreading of misinformation about COVID-19 and vaccines.
We provide a new Multimodal COVID-19 Vaccine Focused Data Repository (MMCoVaR) consisting of images, text, and temporal information.
We combine two news media ranking websites, which allows for better accuracy of the annotation process for the news dataset.
A hybrid stance detection mechanism is proposed for the tweet dataset annotation.
We also provide several statistics and topic analyses on the news dataset and tweet dataset. 
We propose a novel baseline modular architecture for fake news classification using the Longformer. 
In addition, several benchmark performances of misinformation detection on news and tweet datasets are provided.
We expect to continue to add to this data repository by including:
(1) more COVID-19 vaccine-related news articles; 
(2) social media data from more platforms like Sina Weibo (China) and Reddit (U.S.)

\bibliographystyle{unsrt}  
\bibliography{refs}

\end{document}